# Enhanced Quantum Key Distribution using Hybrid Channels and Natural Random Numbers

Hemant Rana, Nitin Verma*

*Amity School of Engineering and Technology, New Delhi*

hemant.rana19979@gmail.com, vermanitin1998@gmail.com

**Abstract**

Since the introduction of quantum computation by Richard Feynman in 1982, Quantum computation has shown exemplary results in various applications of computer science including unstructured database search, factorization, molecular simulations to name a few. Some of the recent developments include quantum machine learning, quantum neural networks, quantum walks on graphs, fault tolerant scalable quantum computers using error correction codes etc. One of the crucial modern applications of quantum information is quantum cryptography and secure key distribution over quantum channels which have several advantages over classical channels, especially detection of eavesdropping. Based on such properties of quantum systems and quantum channels, In this paper we propose three secure key distribution protocols based on a blend of classical and quantum channels. Also the proposed protocols exploits the property of quantum computers to generate natural random numbers that can be easily transmitted using a single qubit over a quantum channel and can be used for distributing keys to the involved parties in a communication network.

## 1. Introduction

Quantum computation is a non-classical computing paradigm proposed by Richard Feynman in 1982. Since then a rigorous development has been observed in the field. Some of the prominent early works include Polynomial time prime factorization by Peter Shor et. al. [1] in 1994 and two years later a quantum algorithm for Unstructured Database search with O $(\sqrt{N})$ complexity by L.K. Grover [2]. With the advent of quantum computing two complexity classes were introduced namely BQP (bounded-error quantum polynomial time) and QMA (Quantum Merlin Arthur) for classifying problems based on quantum computers. All the problems that can be solved by a quantum computer in a polynomial time are classified under BQP class whereas if the problem's solution is verifiable in a polynomial time, it is classified under QMA. Some of the recent popular research topics in quantum computers are Quantum Cryptography, Quantum Image Processing, Quantum Neural Networks, Quantum Machine Learning and Quantum Key Distribution.

Some of the recent works include Big data clustering [3], Representing images in a quantum system using Quantum States and Normalized Amplitudes [4], Training Quantum Neural Networks [5], Quantum algorithm for training gaussian processes [6] and Practical implementation of Cancer detection using Quantum Neural Networks [7]. Quantum cryptography (QC) is an advanced approach for classical cryptography. Quantum cryptography uses a quantum bit also known as a qubit. QC is further divided into 2 broad branches [8]:

a. Quantum Key Distribution (QKD)

b. Quantum Bit Commitment (QBC)

The primary principle of quantum computing, quantum entanglement, works in the core of QKD. The security of quantum key distribution restricts the eavesdropper to listen to the channel and copy the bits. Due to non-cloning property of the qubits and quantum mechanics, the channel remains secure and immune [9]. Due to the advancement in quantum computation, many algorithms such as RSA, ECC and many more will be rendered insecure in the future [10].

In this paper we are proposing three key distribution algorithms based on a combination of quantum and classical channels and the capability of quantum computers to generate natural random numbers. Fig.1 represents the quantum channel and the classical channel. The quantum channel is used to transfer the qubits and the classical channel is used to transfer the classical bits.

The paper is organized as follows. Section 2 gives the overview of the previous related work done in the field of quantum cryptography. Section 3 describes the background details necessary to understand the basics of quantum mechanics and quantum cryptography. Section 4 describes the proposed protocols. Section 5 gives the detailed security analysis of out proposed protocols. Section 6 describes conclusion along with the applications of quantum cryptography.

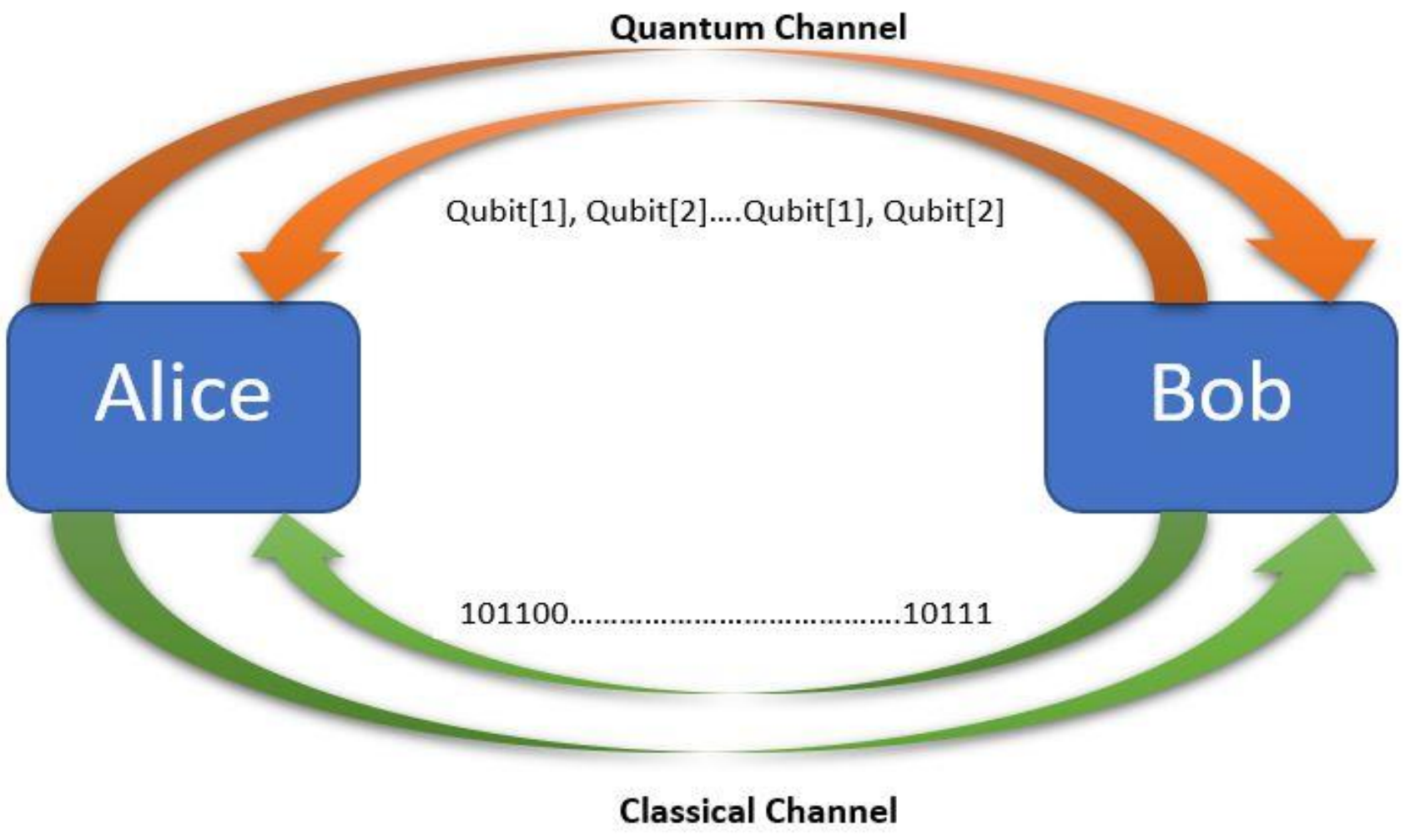

Fig. 1 Quantum channel and Classical channel

## 2. Literature Review

Quantum cryptography has its roots in the concept of quantum money. This concept of quantum money was first proposed in 1969 by Weisner[10]. However due to limitations in technological advancement, his work wasn't published till 1983.

In 1984, Bennett and Brassard proposed the first ever practical QKD protocol, BB84 [11]. BB84 was based on the polarization of the photon. The polarization of the photons was used to establish a key, which was then used for encipherment and decipherment of the plain text. BB84 also used 2 channels for communication- quantum channel to transfer the qubits and the other, classical channel which was used to transfer the data. After the BB84 protocol was proposed, great effort was put in to increase the efficiency as well as the security of QKD algorithms. Ekert then in 1991 proposed another protocol which was based on Bell's theorem [12]. Ekert's protocol used an EPR pair which is essentially a pair of quantum bits. Bennett then in 1992 proposed another algorithm which was an improvement of the BB84 protocol [13]. This new protocol employed two non-orthogonal states. This new protocol was found to be an improvement both in terms of simplicity as well as efficiency. In 1995, Huttner et al. [14] and subsequently in 1998, Brub [15] proposed their own algorithms which were based on the fundamental principles of quantum computing.

In 2007, Mohsen S. and Hooshang A. [16] presented a modification to BB84 protocol. The modification claimed to increase the efficiency of the BB84 algorithm logically. Although they increased the length of the quantum key, the main drawback was the longer time of key generation. In 2009, Hui Q. and Xiao Ch. [17] then proposed another algorithm. They employed BB84 in a depolarization channel as the QKD model. In 2010, Sufyan T. and Omer K. [18] proposed another algorithm. This algorithm integrated QKD with the techniques of that of unconditional secure authentication of classical cryptography. They were able to reduce the authentication cost, but this also reduced the efficiency of the algorithm. In 2012, Marcin N. and Andrzej R. [19], proposed a new concept of entropy. They divided the security of QKD into levels, the basic or the lower level and the advanced or the higher level. Based upon the requirements of the end user the suitable level of security could be employed. Amrin M. et al. [20] in 2013 proposed another algorithm which was able to handle the issue of Man-In-The-Middle form of attack. They proposed two types of key, computational key and information theoretic key, both constitute a hybrid key, and further communication was then based on it. The main disadvantage of this technique was scalability and offline key establishment.

Another key technology for privacy protection in cryptology is quantum oblivious transfer (OQT). The OQT protocol was first put forward by Cr´epeau in the year 1994 [21]. Since then many algorithms based in quantum oblivious transfer have been proposed [22, 23].

Quantum authentication (QA) was first proposed in 2001 Curty and Santos [24]. Since then many QA algorithms have been proposed [25,26].

## 3. Background

Some of the commonly used terms in the field of quantum mechanics and quantum cryptography are:

### *3.1 Quantum bit or a qubit*

Just like a classical bit i.e. 0 or 1 is used to represent information in a classical computer, a qubit is used to represent information in a quantum computer [11]. This qubit is quite different from classical bits. There are only two possible states in classical representation-0 or 1. At a particular given time, the bit can be either a 0 or a 1. However, a qubit can hold 0, 1 or both at the same time.

### *3.2 Basic principles of quantum mechanics*

There are 3 basic principles of quantum mechanics:

i) *Quantum superposition*

Superposition is one of the most crucial properties that differentiates a qubit from a classical bit. The principle of superposition states that a qubit can be represented by |0>, |1> or a combination of these two at any particular instance of time. It is also important to note that if any two valid quantum states are in superposition, then their linear combination will also be a valid quantum state. This means that if the probability of representing the |0> state is 'p' then the probability of representing the |1> state is '1-p', according to the probability theorem [12].

The superposition can be represented in mathematical equation as:

$$\alpha|0\rangle + \beta|1\rangle = |\Psi\rangle \qquad (1)$$

In order for this equation to be normalized, the equation (1) must follow $\alpha^2 + \beta^2 = 1$.

The states |0>, |1> and |Ψ> which is a linear combination of $\alpha|0\rangle + \beta|1\rangle$, are considered as a single qubit state.

In order to visualize this qubit a representation known as Bloch Sphere representation is used. This representation uses three axes namely *x axis, y axis* and *z axis.* The qubit can be represented accurately using this representation.

ii) *Quantum entanglement*

Quantum entanglement is a special connection that exists between two quantum bits. The connection between multiple quanta is so strong that irrespective of the distance between the qubits, the measurement of one qubit will determine the measurement of the other qubit. This entanglement can also be achieved between two qubits, if one qubit is passed through a Hadamard Gate or H gate and then both the qubits are passed through a Controlled X or Controlled NOT or CX gate.

iii) *Quantum interference*

Quantum interference is somewhat similar to wave interference. This interference is due to a phenomenon known as phase. Similar to wave interference, if two qubits are in phase, their amplitude is added together, and if they are out of phase, their amplitudes cancel each other's effect.

### *3.3 Quantum information properties*

There are 3 basic properties for quantum information:

i) *Uncertainty Principle*

The main principle of Heisenberg's Uncertainty Principle is that it is impossible to determine the accurate position and velocity of a particle either theoretically or practically. This theorem can be represented mathematically as:

$$\Delta x \Delta p \geq (h/4*\Pi) \qquad (2)$$

where, $\Delta x$ = position

$\Delta p$ = momentum

h = Planck's constant

ii) *No-clone theorem*

According to this theorem, a qubit that is being transmitted between two parties involved in communication, cannot be copied or cloned by a third party that tries to intercept the communication by inserting itself in the communication channel. This property along with quantum entanglement have been proved to be quite useful.

iii) *Quantum Teleportation*

Quantum teleportation enables a sender to send the quantum information to the receiver using quantum entanglement, given the entangled qubits have been shared between both the involved parties. An example of quantum information is the precise state or measurement of a photon or an atom.

### *3.4 Basic gates used in quantum computing*

There are four basic gates that are most widely used in quantum mechanics:

i) *Hadamard Gate (H-Gate)*

The H-Gate is a rotational gate. It is useful to rotate the state |0> to |+> and |1> to |->. This gate is widely used to achieve quantum superposition. H-Gate operates on a single qubit. The equation of H-Gate is:

$$H = (|0\rangle + |1\rangle) / \sqrt{2} \qquad (3)$$

ii) *Controlled NOT Gate (CX-Gate)*

The CX-Gate works on a pair of qubits. One qubit is known as the controlled bit and the other as target bit. Whenever control qubit is |1>, the CX-Gate performs the NOT operation on the target qubit, irrespective of its initial state. One compelling property of CX-Gate is that if the controlled bit is in superposition and then it is passed through a CX-Gate with the target bit, the CX-Gate achieves entanglement within the two.

iii) *Pauli Z Gate (Z-Gate)*

Pauli Z Gate is a single input gate. This gate has the ability of performing the flip operation. It flips the |+> to |->. This is achieved by rotating the qubit by $\pi$ radians along the *z-axis* of the Bloch Sphere representation. When the qubit is in state |0> the qubit is passed as it is, however if the qubit is in state |1>, the qubit is converted into -|1> state.

iv) *SWAP Gate*

The SWAP Gate is the simplest gate. It is used to perform the SWAP operation on the qubits. It is a two-input gate. SWAP-Gate simply swaps the state of the two qubits involved.

### *3.5 EPR Pair and Bell State*

Bell State is a maximally entangled state involving two qubits such that the probability of measuring either of the two qubits is equal [13]. Since according to the principle of probability, the sum of all the probability should be equal to 1, therefore the probability of measuring either of the two qubits of Bell state comes out to be half or 0.5. A two-bit quantum system can exist in any one of the following Bell State:

$$|\Phi_{+}\rangle = ( 1 / \sqrt{2} ) * ( |00\rangle + |11\rangle )$$

$$|\Phi_{-}\rangle = ( 1 / \sqrt{2} ) * ( |00\rangle - |11\rangle )$$

$$|\psi_{+}\rangle = ( 1 / \sqrt{2} ) * ( |01\rangle + |10\rangle )$$

$$|\psi_{-}\rangle = ( 1 / \sqrt{2} ) * ( |01\rangle - |10\rangle ) \quad (4)$$

An EPR pair, or Einstein-Podolsky-Rosen Pair is a pair of quantum bits, that are maximally entangled and are in Bell state together.

## 4. Proposed Algorithm

Since the protocols make use of random numbers which can be efficiently generated by quantum computers, hence the following function is used for generating random numbers using a quantum bit.

*RandomNumber* (qubit Q, integer n):

```
from 1 to n:
        Measure(Q)
p := pr (desired state of Q)
q := pr (undesired state of Q)
```

$$\text{randomNumber} = \frac{\log(p)}{2(1+\log(q))}$$

```
return randomNumber
```

where pr($x$)stands for probability of event $x$.

The aforementioned function has been used for generating a random number lying between 0 and 1 because the stated algorithm has a less complexity in the context of both classical and quantum computing. Also, the function can generate a random number to several decimal places for smaller values of n.

*4.1 Notations and Guidelines*

- $\oplus$ represents bitwise XOR operation.
- $||$ represents concatenation.
- $U(x)$ represents a unitary operation on qubit $x$.
- All the qubits are initialized and measured in a common state. Measure function outputs the probability of occurrence of the desirable state.

*Protocol 1:*

| ALICE | BOB |
|---|---|
| $R_A$ = RandomNumber $(Q_A, n)$<br><br>Consider two entangled qubits $(Q_{A_1}, Q_{A_2})$<br><br>$U(Q_{A_1}, Q_{A_2})$<br><br>**Transmit** $Q_{A_2}$ to BOB through a quantum channel.<br><br>$M_1$ = measure $(Q_{A_1})$<br>$M_3$ = measure $(Q_{B_2})$<br><br>$M_{13} = M_1 \oplus M_3$<br><br>**Transmit** $M_{13}$ to BOB using classical channel.<br><br>$M_{1234} = M_{13} \mid\mid M_{24}$<br>$MR_A = M_{1234} \oplus R_A$<br><br>**Transmit** $MR_A$ to BOB using classical channel.<br><br>$MR_{AB} = MR_A \oplus MR_B$<br><br>Consider two entangled qubits $(Q_{A_3}, Q_{A_4})$ | $R_B$ = RandomNumber $(Q_B, n)$<br><br>Consider two entangled qubits $(Q_{B_1}, Q_{B_2})$<br><br>$U(Q_{B_1}, Q_{B_2})$<br><br>**Transmit** $Q_{B_2}$ to ALICE through a quantum channel.<br><br>$M_2$ = measure $(Q_{B_1})$<br>$M_4$ = measure $(Q_{A_2})$<br><br>$M_{24} = M_2 \oplus M_4$<br><br>**Transmit** $M_{24}$ to ALICE using classical channel.<br><br>$M_{1234} = M_{13} \mid\mid M_{24}$<br>$MR_B = M_{1234} \oplus R_B$<br><br>**Transmit** $MR_B$ to ALICE using classical channel.<br><br>$MR_{AB} = MR_A \oplus MR_B$<br><br>Consider two entangled qubits $(Q_{B_3}, Q_{B_4})$ |
| $U(Q_{A_3}, Q_{A_4})$<br><br>**Transmit** $Q_{A_4}$ to BOB through a quantum channel.<br><br>$M_5$ = measure ($Q_{A_3}$)<br>$M_6$ = measure ($Q_{B_4}$)<br><br>***Key*** = hash ( $M_5 \|\| M_6 \|\| MR_{AB}$ ) | $U( Q_{B_3}, Q_{B_4})$<br><br>**Transmit** $Q_{B_4}$ to ALICE through a quantum channel.<br><br>$M_7$ = measure ($Q_{B_3}$)<br>$M_8$ = measure ($Q_{A_4}$)<br><br>***Key*** = hash ( $M_8 \|\| M_7 \|\| MR_{AB}$ ) |

*Protocol 2:*

| ALICE | BOB |
|---|---|
| $R_A$ = RandomNumber $(Q_A, n)$<br><br>Consider two entangled qubits $(Q_{A_1}, Q_{A_2})$<br><br>$\theta = cos^{-1}(R_A)$<br><br>Consider a qubit $Q_{A_3}$<br><br>$Z_\theta( Q_{A_3} )$<br><br>$U(Q_{A_1}, Q_{A_2})$<br><br>**Transmit** $Q_{A_2}, Q_{A_3}$ to BOB using quantum channel.<br><br>$M_2 = measure(Q_{B_2})$<br>$M_4 = measure(Q_{A_1})$<br><br>$R_{AB} = ((M_2 \oplus M_4) \|\|R_B) \oplus R_A) * R_B$<br><br>***Key*** = hash ($R_{AB}$) | $R_B$ = RandomNumber $(Q_B, n)$<br><br>Consider two entangled qubits $(Q_{B_1}, Q_{B_2})$<br><br>$\theta = cos^{-1}(R_B)$<br><br>Consider a qubit $Q_{B_3}$<br><br>$Z_\theta( Q_{B_3} )$<br><br>$U(Q_{B_1}, Q_{B_2})$<br><br>**Transmit** $Q_{B_2}, Q_{B_3}$ to ALICE using quantum channel.<br><br>$M_1 = measure(Q_{A_2})$<br>$M_3 = measure(Q_{B_1})$<br><br>$R_{AB} = ((M_3 \oplus M_1) \|\|R_B) \oplus R_A) * R_B$<br><br>***Key*** = hash ($R_{BA}$) |

*Protocol 3:*

| ALICE | BOB |
|---|---|
| Consider two entangled qubits $(Q_{A_1}, Q_{A_2})$<br><br>$U(Q_{A_1}, Q_{A_2})$<br><br>**Transmit** $Q_{A_2}$ to BOB using quantum channel.<br><br>$M_1$ = measure $(Q_{A_1})$<br>$M_3$ = measure $(Q_{B_2})$<br><br>$C = M_1 \oplus M_3$<br><br>$R_{A_1} = randomNumber(Q_{RA_1}, n_1)$<br>$R_{A_2} = randomNumber(Q_{RA_2}, n_2)$<br><br>$Y = [R_{A1} \| C \| R_{A2}] \oplus C$<br><br>**Transmit** Y to BOB using classical channel.<br><br>$K = (R_{A_1} \oplus R_{B_1}) \| (R_{A_2} \oplus R_{B_2})$<br><br>***Key*** = hash $(K \| C)$ | Consider two entangled qubits $(Q_{B_1}, Q_{B_2})$<br><br>$U(Q_{B_1}, Q_{B_2})$<br><br>**Transmit** $Q_{B_2}$ to ALICE using quantum channel.<br><br>$M_2$ = measure $(Q_{A_1})$<br>$M_4$ = measure $(Q_{B_2})$<br><br>$C = M_2 \oplus M_4$<br><br>$R_{B_1} = randomNumber(Q_{RB_1}, n_3)$<br>$R_{B_2} = randomNumber(Q_{RB_2}, n_4)$<br><br>$Z = [R_{A1} \| C \| R_{A2}] \oplus C$<br><br>**Transmit** Z to ALICE using classical channel.<br><br>$K = (R_{A_1} \oplus R_{B_1}) \| (R_{A_2} \oplus R_{B_2})$<br><br>***Key*** = hash $(K \| C)$ |

## *4.2 Important features*

- The measurement of the qubits can be done in any basis by the entities involved, i.e. Alice and Bob. This basis has to be selected beforehand. The measurement and selection of the basis solely depends upon the choice of the user. This also helps to create randomness, which aids in enhancement of the security of the protocol.
- The key is generated using a hash function such as MD5 or SHA-256, so even a slight change will produce a distinctive hash and therefore a unique key each time.
- Another advantage of a hash function as a key is that, the output of a hash function is always a fixed size irrespective of the input length.

- The key produced by any of the protocol is a *one-time-key.* Once a session is terminated and a new session is to be established, the entire process has to be repeated. Repeating the entire protocol after each session makes the session immune and protects the Confidentiality and Integrity.
- Since according to no-cloning theorem, state of a qubit cannot be copied by the intercepting entity, this also helps to improve the security of the protocols.

## 5. Security Analysis and Result

Following figures depicts the output for an experiment implemented using IBM Q for various iterations.

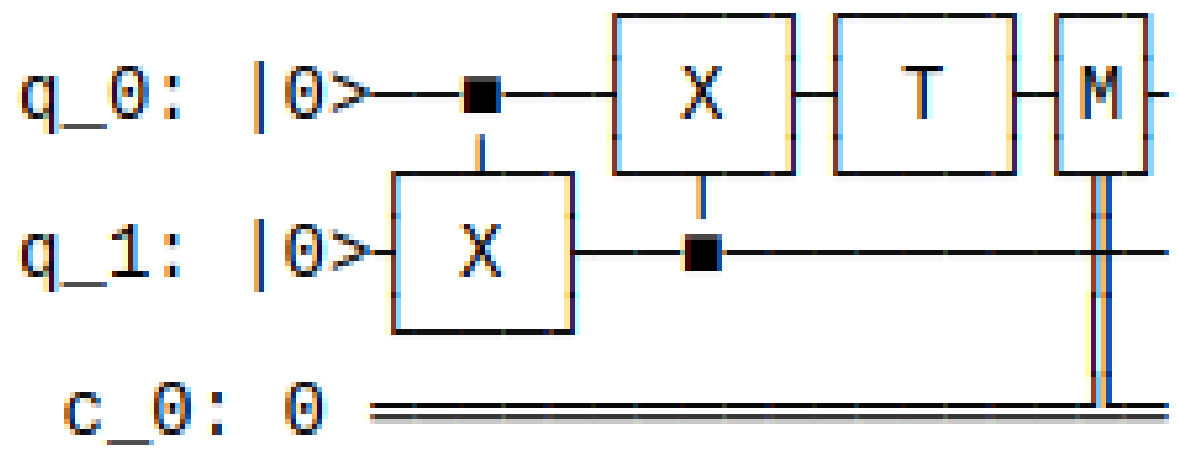

Fig. 2 Circuit for generating random number

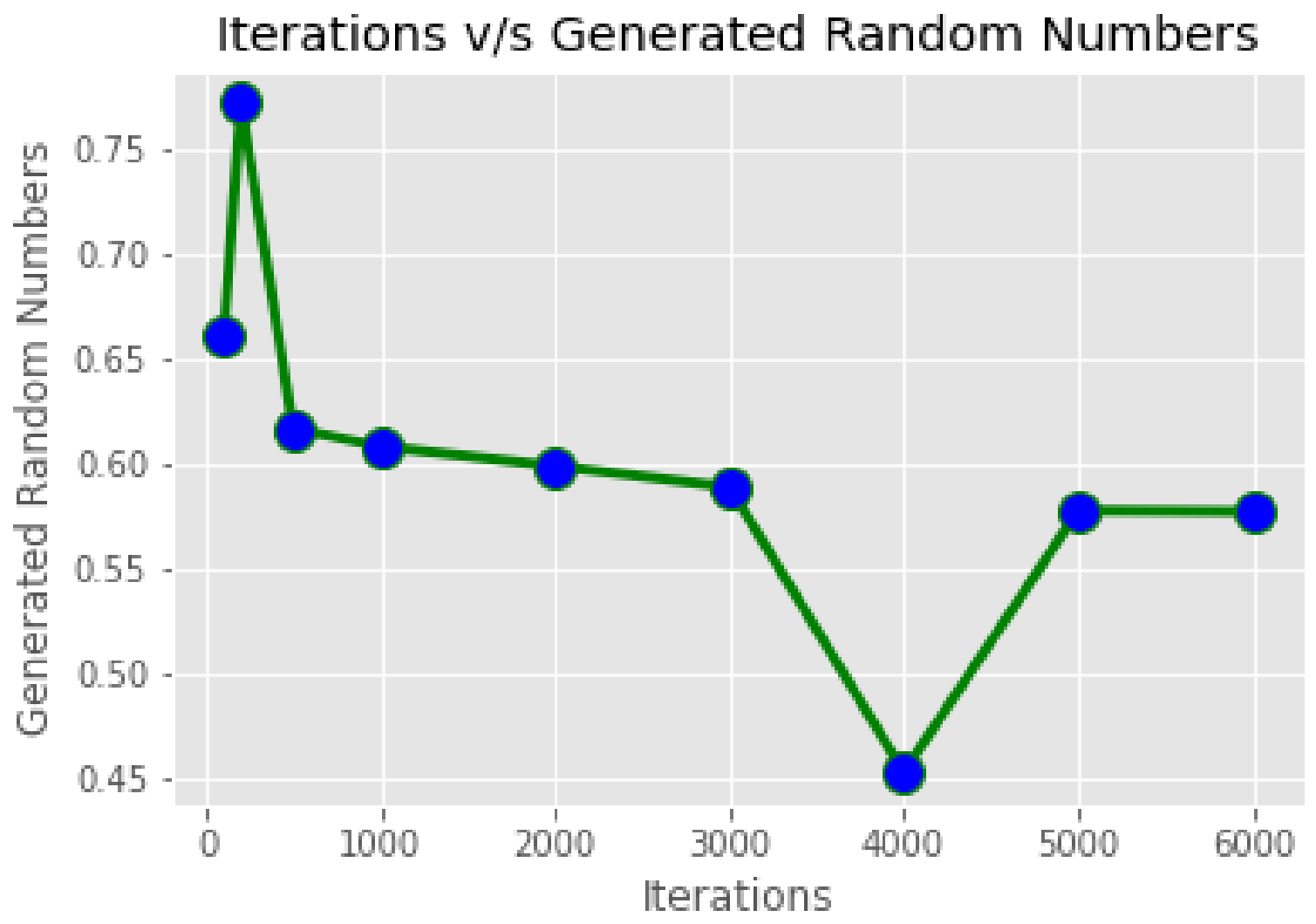

Fig. 3 Plot for n versus generated random number

It can be observed that the proposed protocols are unconditionally secure. Instead of using a single method for key generation, the proposed protocols go through a number of steps after which the key is generated. The key generated is then used for symmetric cryptography which is more efficient as compared to asymmetric cryptography. The proposed protocols also use the XOR and concatenation operations. These operations produce minimal overhead while maintaining the overall security of the system. The use of hash function to generate the key also makes the protocols immune and secure.

It is also evident that even if the eavesdropper is listening on the classical channel and is able to intercept the data that is exchanged between Alice and Bob, the eavesdropper will not be able to generate the actual key which is used to encrypt or decrypt the message.

The ability of quantum computers to generate truly random numbers is also being used. Since quantum numbers can generate true random numbers, this property also aids in enhancing the overall security of the system.

The use of generated keys as one-time-key also helps in maintaining secure transmission and reception. A key used once cannot be used again. The operations with minimal overhead i.e. concatenation and XOR, helps in maintaining efficiency. The true random generator maintains security by making sure that a new random number is generated each time without any correlation with the previously generated number.

Another notable feature of the protocols is that, they are generating the keys without the involvement of any kind of third-party dependency, which may be involved in the process of key generation. This also enhances the security as the cryptographic key is known only to the parties involved in the message transfer.
The proposed algorithms are more efficient as compared to existing algorithms as they required n qubits. This produced an unnecessary overhead of encoding and transmitting each qubit over the network, whereas in the proposed algorithms the number of qubits are limited to a single digit, irrespective of their length.

The schemes described are also more flexible and scalable. Additionally, it also provides mutual authentication and added security while preserving data integrity.

## 6. Conclusion

Cryptography plays an important role in maintaining the Confidentiality and Integrity of any system. The data that is being exchanged between the sender and the receiver is always private irrespective of its nature. In the future quantum computing will play a huge role in terms of computational complexity. Hence quantum cryptography will play an integral part in maintaining the privacy of the data being transmitted. Some of the applications of quantum cryptography are:

i) Secure communications between space and earth
ii) Smart Power Grid
iii) Quantum Internet
iv) Voting with high security

In this paper we proposed three new quantum key distribution protocols which are based on the following properties:

i) Ability to generate natural true random numbers.
ii) Encoding a qubit with any value in the range [0,1] .
iii) No cloning theorem of a qubit.
iv) Quantum Entanglement.

The algorithms provide high security by providing minimum overhead, higher efficiency, higher flexibility and higher scalability.